# Calculation of the Electron Spin Relaxation Times in InSb and InAs by the Projection-Reduction Method

Nam Lyong Kang

*Department of Nanomechatronics Engineering,*
*Pusan National University, Miryang 627-706, Republic of Korea*

Abstract

The electron spin relaxation times in a system of electrons interacting with piezoelectric phonons mediated through spin-orbit interactions were calculated using the formula derived from the projection-reduction method. The results showed that the temperature and magnetic field dependence of the relaxation times in InSb and InAs were similar. The piezoelectric materials constants obtained by a comparison with the reported experimental result were $P_{\text{pe}} = 4.0 \times 10^{21}$ eV/m for InSb and $P_{\text{pe}} = 6.7 \times 10^{21}$ eV/m for InAs. The relaxation of the electron spin can be explained by the Elliot-Yafet process at a high field limit.

After the studies of the spin dynamics by Lampel [1] and Parsons [2], numerous experimental and theoretical investigations have been carried out on spintronics in semiconductors [3-11]. Preserving the spin information over a practical device length and time scale is important for realizing spin-based electronic and optoelectronic devices. Therefore, understanding the spin relaxation and dephasing mechanisms in semiconductors is of great importance for the practical use of such devices.

For the materials used widely in spintronics, such as III-V and II-VI compounds, the relevant spin relaxation and dephasing mechanisms include the Elliot-Yafet (EY) [12,13], D'yakonov-Perel' (DP) [14], Bir-Aronov-Pikus (BAP) [15,16], and g-tensor inhogenity [17] mechanisms. Spin relaxation and dephasing in semiconductors are a consequence of the coaction of these mechanisms. Therefore, it is important to determine the dominant spin relaxation mechanism under a range of conditions, such as temperature, external field and doping density. The BAP mechanism is ineffective due to the lack of holes and the EY mechanism is less important than the DP mechanism in n-type semiconductors [18], such as InSb and InAs, which are the prospective materials for both high speed electronic and spintronic devices because of their high electron effective g-value, small electron effective mass, and high mobility.

On the other hand, it is important to understand how the phonon and electron distribution functions are included because the temperature dependence of the electron spin relaxation time might be affected by the distribution functions. In this paper, the electron spin relaxation times in a system of electrons interacting with piezoelectric deformation phonons through phonon-modulated spin-orbit interaction are calculated using the formula obtained from Kang-Choi's projection-reduction (KCPR) method [19]. The formula satisfies the "population criterion", which states that the electron and phonon distribution functions should be combined in multiplicative forms because the electrons and phonons belong to different categories in a quantum-statistical classification, and the formula can be interpreted by diagram, which can give an intuition for the quantum dynamics of electrons in solids [20]. The temperature, magnetic field, and electron density dependence of the relaxation times in InSb and InAs are obtained by a comparison with the experimental data reported by Litvinenko et al. [21].

Using the KCPR method and considering the Lorentzian approximation for weak electron-phonon interactions, the electron spin relaxation time ($T_1$) for the EY process can be

expressed as [19,20]

$$\frac{1}{T_1} = \frac{2\pi}{\hbar(f_{\alpha-} - f_{\alpha+})} \sum_{\gamma,q}$$
$$\times \{W(\alpha-,\gamma+)2|[l_z(q)]_{\gamma+,\alpha+}|^2 + 2|[l_z(q)]_{\alpha-,\gamma-}|^2 W(\gamma-,\alpha+)$$
$$+|[l^+(q)]_{\alpha-,\gamma+}|^2 W(\gamma+,\alpha+) + W(\alpha-,\gamma-)|[l^+(q)]_{\gamma-,\alpha+}|^2\}. \quad (1)$$

where $f_{\alpha s}$ is the Fermi distribution function for an electron with energy $E_{\alpha s}$, where $s = +(-)$ for an up (down) spin. The energy eigenvalue under a static magnetic field $B$ applied in the $z-$ direction is given as $E_{\alpha s} = (n_\alpha + 1/2)\hbar\omega_c + \hbar^2 k_{z\alpha}^2/2m_e + g\mu_B B s/2$, where $n_\alpha = 0,1,2,\cdots$, $\omega_c$ is the cyclotron frequency, $k_{z\alpha}$ is the $z-$component of the electron wave vector, $g$ is the electron g-factor, and $\mu_B$ is the Bohr magneton. In Eq. (1), $[l_z(q)]_{\alpha s,\beta s'}$ and $[l^+(q)]_{\alpha s,\beta s'}$ are the interaction coupling factors, which are given later, and the transition factor, $W(\alpha s, \beta s')$, is defined as

$$W(\alpha s, \beta s') \equiv \delta(\hbar\omega + E_{\alpha s} - E_{\beta s'} - \hbar\omega_q)P_+(\alpha s, \beta s')$$
$$+\delta(\hbar\omega + E_{\alpha s} - E_{\beta s'} + \hbar\omega_q)P_-(\alpha s, \beta s'). \quad (2)$$

Here $\delta(x)$ denotes the Dirac delta function, $\omega$ is the frequency of an incident electromagnetic wave, and the population factors, $P_\pm(\alpha s, \beta s')$, are defined as

$$P_+(\alpha s, \beta s') \equiv (1 + N_q)f_{\alpha s}(1 - f_{\beta s'}) - N_q f_{\beta s'}(1 - f_{\alpha s}) \quad (3)$$
$$P_-(\alpha s, \beta s') \equiv N_q f_{\alpha s}(1 - f_{\beta s'}) - (1 + N_q)f_{\beta s'}(1 - f_{\alpha s}), \quad (4)$$

where $N_q$ is the Planck distribution function for phonons with an energy $\hbar\omega_q$. Eqs. (3) and (4) satisfy the population criterion.

The electron spin relaxation times in InAs and InSb are calculated using Eq. (1) for $n_\alpha = 0$ at the subband edge ($k_{z\alpha} = 0$) in the quantum limit. Hence, only the cases $n_\gamma = 0$ and $n_\gamma = 1$ need to be considered in Eq. (1). In these cases, $|[l_z(q)]_{\alpha\beta}|^2$ and $|[l^+(q)]_{\alpha\beta}|^2$ are given as follows [see Eqs. (28) and (29) in Ref. 19]: $|[l_z(q)]_{\alpha\gamma}|^2 = (l_a^2 \hbar D_q^2 eBte^{-t}q_\perp^2/\sqrt{2})$ and $|[l^+(q)]_{\alpha\gamma}|^2 = \sqrt{2}\hbar eB l_a^2 D_q^2 q_z^2 t e^{-t}$ for $n_\gamma = 0$, $|[l_z(q)]_{\alpha\gamma}|^2 = [l_a^2 \hbar D_q^2 eBe^{-t}(1-t)^2 q_\perp^2/\sqrt{2}]$ and $|[l^+(q)]_{\alpha\gamma}|^2 = \sqrt{2}\hbar eB l_a^2 D_q^2 q_z^2 e^{-t}(1-t)^2$ for $n_\gamma = 1$, where $\delta_{k_{y\gamma},k_{y\alpha}-q_y}\delta_{k_{z\gamma}-q_z}$ are omitted, $l_a \equiv \hbar/4m_e^2 c^2$, $t \equiv \hbar q_\perp^2/2eB$, and $q_\perp^2 \equiv q_x^2 + q_y^2$. The electron-phonon coupling factor, $D_q$, is given as $D_q \equiv \sqrt{\hbar/2\rho_m \Omega_0 \omega_q} P_{pe}^2 q^2/(q^2 + q_d^2)$ for the piezoelectric deformation potential [20]. Here, $q_\mathrm{d} = \sqrt{n_e e^2/\varepsilon\varepsilon_0 k_\mathrm{B} T}$ is the reciprocal of the Debye screening length, where $n_\mathrm{e}$ is the number density of electrons and $\varepsilon$ is the static dielectric constant. Note that the proportional constant (piezoelectric material constant), $P_\mathrm{pe}$, is used as a fitting parameter and only affects the magnitude of the spin relaxation time, i.e. they do not affect the temperature dependence of the spin relaxation time because they are not contained in the distribution functions for electrons and phonons.

Figures 1 and 2 show the magnetic field dependence of the electron spin relaxation times by the piezoelectric deformation potential scattering in InSb and InAs for $n_\mathrm{e} = 1 \times 10^{21}$ m$^{-3}$ at 100 K. Fitting the present theoretical results to the experimental result reported by Litvinenko et al. [21] yielded $P_\mathrm{pe} = 4.0 \times 10^{21}$ eV/m for InSb and $P_{pe} = 6.7 \times 10^{21}$ eV/m for InAs. The theory was fitted to the experiment at a high field because the quantum limit was considered. The discrepancies at low fields may be corrected if the transitions between the high order Landau levels ($n_\alpha \geq 1$) and the DP mechanism are considered. The Bohr magneton, $\mu_\mathrm{B} = e\hbar/2m_\mathrm{e}$, for InSb is larger than that of InAs because $m_\mathrm{e}(\mathrm{InAs}) = 0.026\, m_0 > m_\mathrm{e}(\mathrm{InSb}) = 0.0135\, m_0$, where $m_0$ is the free electron mass. Therefore, Zeeman splitting in InSb is larger than that in InAs. Consequently, the scattering effect (or relaxation rate) in InAs is smaller than in InSb because the electrons in InAs are scattered by the phonons with low energy. As a result, the relaxation time in InAs is larger than that in InSb because the relaxation time is proportional to the inverse of the relaxation rate.

Figures 3 and 4 show the temperature ($T$) dependence of the electron spin relaxation times ($T_1$) for different electron densities at $B = 1$ T. The relaxation times decrease with increasing temperature as $T_1 \propto T^{-1.1}$ for both InSb and InAs at a low density of $n_\mathrm{e} = 1 \times 10^{21}$ m$^{-3}$, which show reasonable agreement with that of (In,Al)As/AlAs quantum dots, $T_1 \propto T^{-0.9}$ [22]. The relaxation times decrease with increasing temperature and increase with increasing electron density because the effects of phonon scattering increase with increasing number of phonons as the temperature is increased and decrease with increasing screening effect as the electron density is increased. Figure 5 shows that $P_\mathrm{pe}$ affects the relaxation time, but not the temperature dependence.

Table I lists the temperature and magnetic field dependence of the relaxation times in InSb and InAs. The magnetic field dependence of the relaxation times (not shown in the figure) can be obtained in a similar manner using Figs. 1 and 2. The spacing between the energy levels increases with increasing magnetic field, and the energy of the piezoelectric phonons is dependent on the wave vector. Therefore, the relaxation times (rates) decrease (increase) with increasing magnetic field because the electrons are scattered by phonons with high energy. The temperature and magnetic field dependence of the relaxation times in InSb and InAs are similar.

Table I: Temperature and magnetic field dependence of the spin relaxation times in InSb and InAs. The temperature dependence and the magnetic field dependence were obtained at $B = 1$ T and $T = 100$ K, respectively [$T$: temperature, $B$: magnetic field, $T_1$: relaxation time, $n_e$: electron density].

| $n_e [\text{m}^{-3}]$ | InSb | InAs |
|---|---|---|
| $1 \times 10^{21}$ | $T_1 \propto T^{-1.10}$ | $T_1 \propto T^{-1.10}$ |
|  | $T_1 \propto B^{-1.70}$ | $T_1 \propto B^{-1.80}$ |
| $5 \times 10^{21}$ | $T_1 \propto T^{-1.30}$ | $T_1 \propto T^{-1.34}$ |
|  | $T_1 \propto B^{-1.75}$ | $T_1 \propto B^{-1.85}$ |
| $10 \times 10^{21}$ | $T_1 \propto T^{-1.45}$ | $T_1 \propto T^{-1.50}$ |
|  | $T_1 \propto B^{-1.83}$ | $T_1 \propto B^{-1.94}$ |

Thus far, the electron spin relaxation times in InSb and InAs were calculated using the formula obtained using the KCPR method and considering piezoelectric phonon-modulated spin-orbit interactions. $P_{\text{pe}} = 4.0 \times 10^{21}$ eV/m for InSb and $P_{\text{pe}} = 6.7 \times 10^{21}$ eV/m for InAs were obtained by fitting the results to the previously reported experimental result. The temperature and magnetic field dependence of the relaxation times in InSb and InAs were slightly different. The relaxation of the electron spin could be explained by the EY mechanism at a high field limit despite the fact that it is less important than the DP mechanism in n-type semiconductors [18]. This means that how the distribution functions are included in the relaxation time is important at high fields because the temperature dependence of the electron spin relaxation time is caused by the distribution functions. Note that the present result includes the distribution functions properly, i.e., it satisfies the population criterion.

The acoustic strain by pressure gives rise to a macroscopic electric field in a crystal whose lattice lacks inversion symmetry, such as III-V materials. In this paper, it was assumed that the electric field is proportional to the derivative of the atomic displacement. The proportional constants ($P_{\text{pe}}$), and the temperature and magnetic field dependence of the relaxation times for the III-V materials can be obtained using the projection-reduction method used in this paper, and the results can be used to examine the piezoelectricity of III-V materials and process quantum information with spins in semiconductors because $P_{\text{pe}}$ does not affect the dependence. Other scattering mechanisms, such as the DP and BAP mechanisms, will be investigated using the KCPR method in the future.

[1]  G. Lampel, Phys. Rev. Lett. **20**, 491 (1968).
[2]  R. R. Parsons, Phys. Rev. Lett. **23**, 1151 (1969).
[3]  S. Datta and B. Das, Appl. Phys. Lett. **56**, 665 (1990).
[4]  B. E. Kane, Nature **393**, 133 (1998).
[5]  Y. Ohno, Science **281**, 951 (1998).
[6]  J. M. Kikkawa and D.D. Awschalom, Phys. Rev. Lett. **80**, 4313 (1998).
[7]  C. Monroe, Nature **416**, 238 (2002).
[8]  I. Žutic′, J. Fabian, and S. Das Sarma, Rev. Mod. Phys. **76**, 323 (2004).
[9]  I. Appelbaum, B. Huang, and D. J. Monsma, Nature **447**, 295 (2007).
[10]  S. P. Dash, S. Sharma, R. S. Patel, M. P. de Jong, and R. Jansen, Nature **462**, 491 (2009).
[11]  C. Zhao, T. Yan, H. Ni, Z. Niu, and X. Zhang, Appl. Phys. Lett. **102**, 012406 (2013).
[12]  R. J. Elliott, Phys. Rev. **96**, 266 (1954).
[13]  Y. Yafet, in *Solid State Physics*, ed. F. Seitz and D. Turnbull (Academic, New York, 1963), Vol. 14, pp. 1-98.
[14]  M. I. D'yakonov and V.I. Perel', Zh. Eksp. Teor. Fiz. **60**, 1954 (1971).
[15]  G. L. Bir, A.G. Aronov, and G.E. Pikus, Zh. Eksp. Teor. Fiz. **69**, 1382 (1975).
[16]  A. G. Aronov, G.E. Pikus, and A.N. Titkov, Zh. Eksp. Teor. Fiz. f84,1170 (1983).
[17]  M. W. Wu and C.Z. Ning, Eur. Phys. J. B **18**, 373 (2000).
[18]  J. H. Jiang and M.W. Wu, Phys. Rev. B **79**, 125206 (2009).

[19]   N. L. Kang and S. D. Choi, Chin. Phys. B  **22**, 087102 (2014).

[20]   N. L. Kang and S. D. Choi, Jpn. J. Appl. Phys.  **53**, in press (2014).

[21]   K. L. Litvinenko, M. A. Leontiadou, J. Li, S. K. Clowes, M. T. Emeny, T. Ashley, C. R. Pidgeon, L. F. Cohen, and B. N. Murdin, Appl. Phys. Lett.  **96**, 111107 (2010).

[22]   D. Dunker, T. S. Shamirzaev, J. Debus, D. R. Yakovlev, K. S. Zhuravlev, and M. Bayer, Appl. Phys. Lett.  **101**, 142108 (2012).

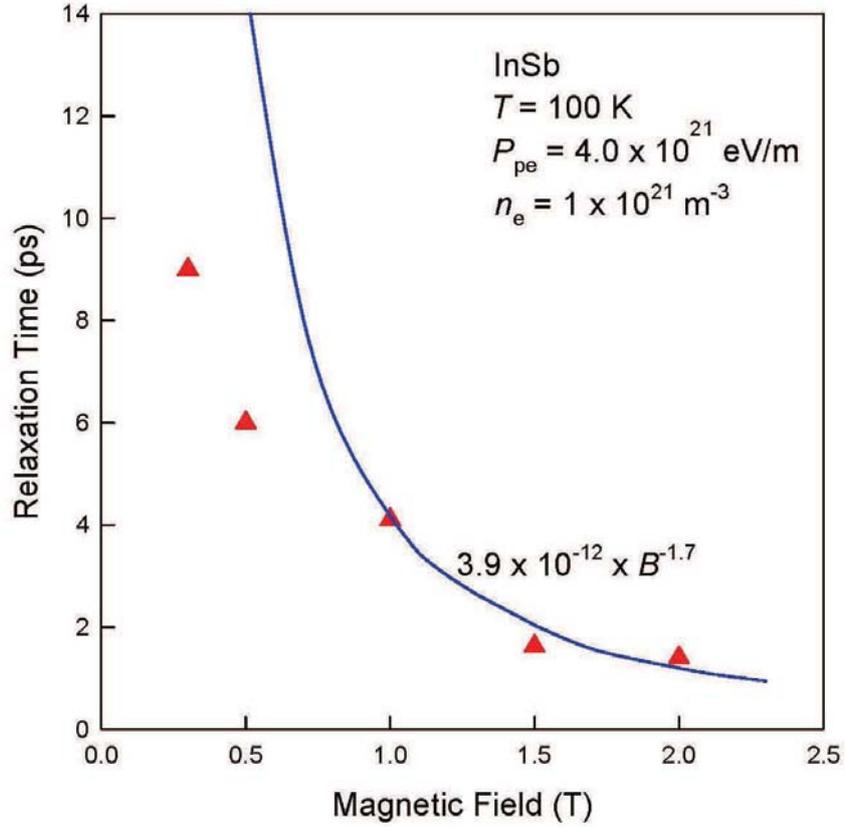

FIG. 1. Magnetic field dependence of the electron spin relaxation time in InSb for $n_e = 1 \times 10^{21} \text{m}^{-3}$ and $P_{pe} = 4.0 \times 10^{21} \text{eV/m}$ at $T = 100\text{K}$. The red triangles and blue solid line are the results reported by Litvinenko et al.[21] and obtained using the present method, respectively.

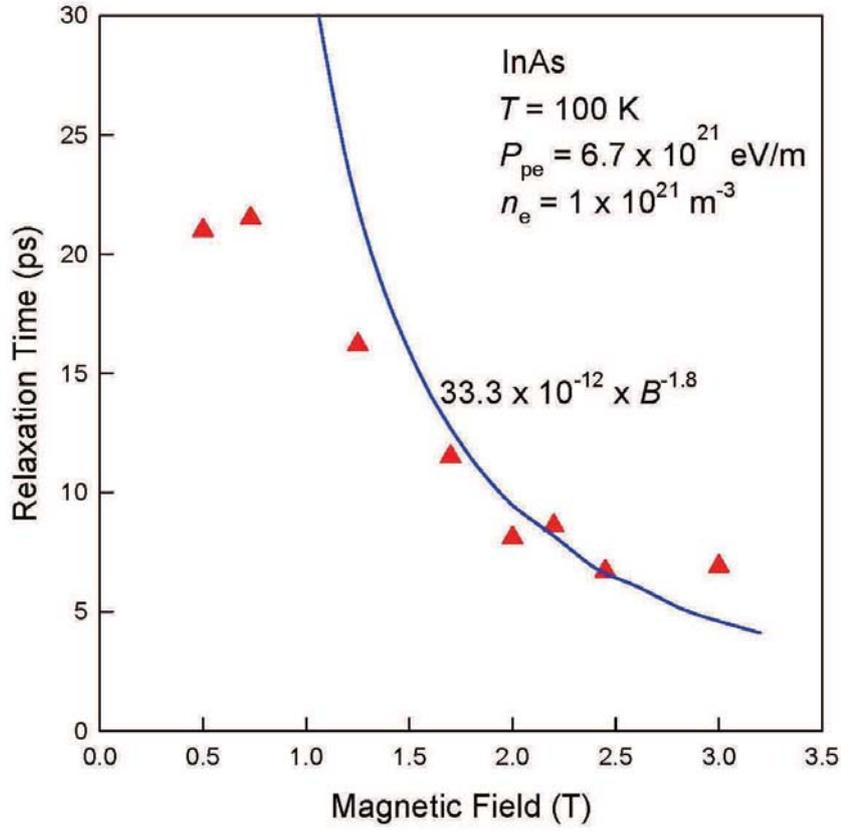

FIG. 2. Magnetic field dependence of the electron spin relaxation time in InAs for $n_e = 1 \times 10^{21} \text{m}^{-3}$ and $P_{pe} = 6.7 \times 10^{21} \text{eV/m}$ at $T = 100\text{K}$. The red triangles and blue solid line are the results reported by Litvinenko et al.[21] and obtained using the present method, respectively.

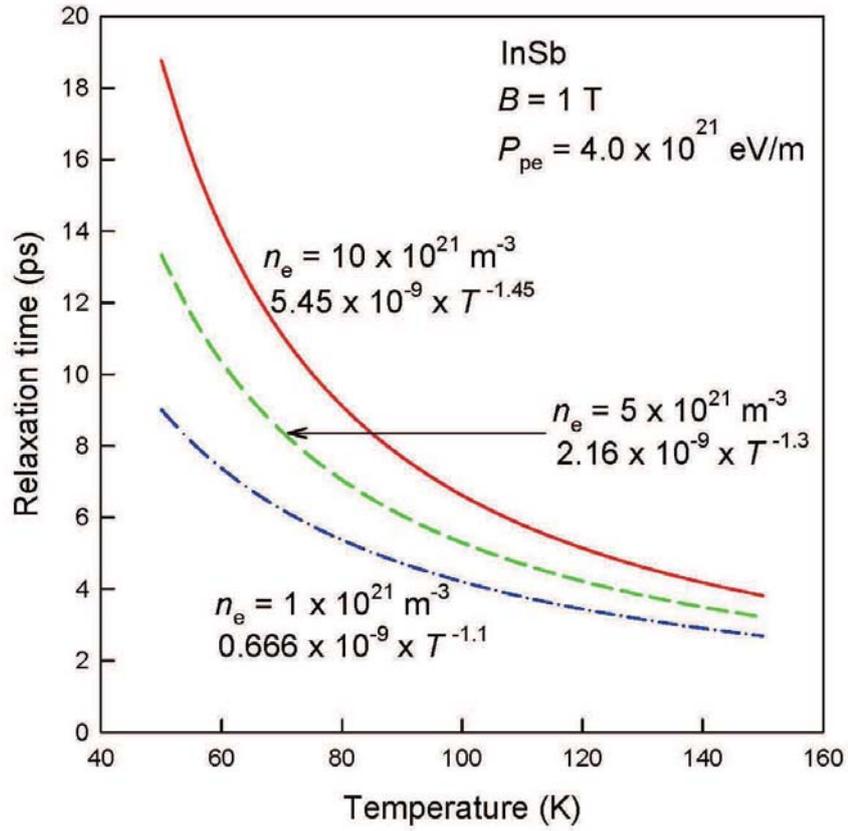

FIG. 3. Temperature dependence of the electron spin relaxation times in InSb for $P_{\text{pe}} = 4.0 \times 10^{21}\,\text{eV/m}$ and various electron densities at $B = 1\,\text{T}$.

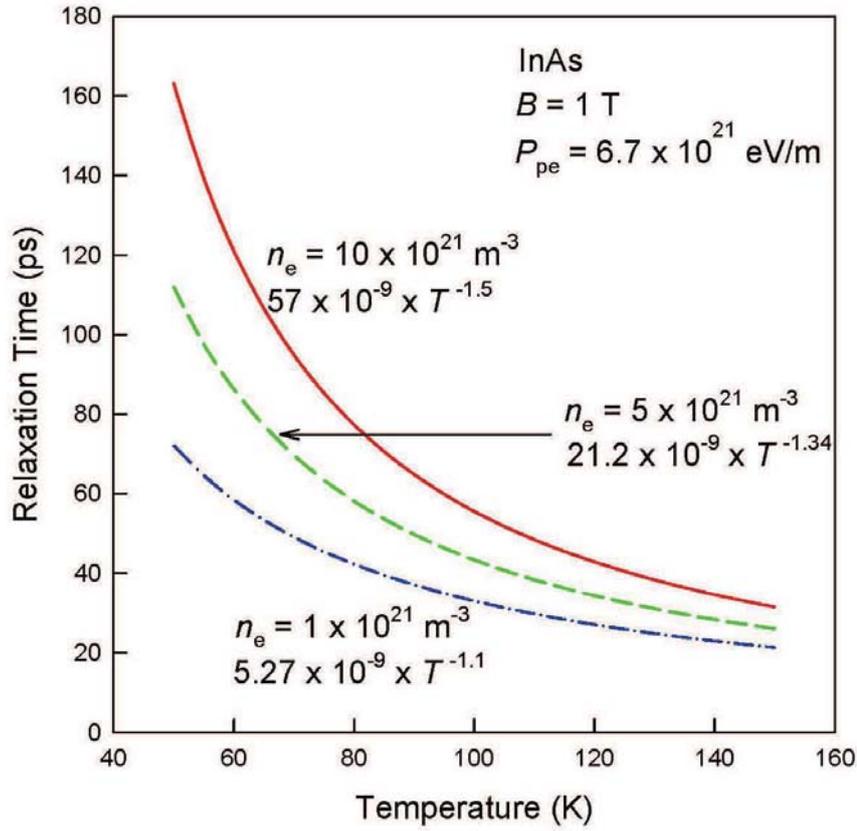

FIG. 4. Temperature dependence of the electron spin relaxation times in InAs for $P_{pe} = 6.7 \times 10^{21}\,\mathrm{eV/m}$ and various electron densities at $B = 1\,\mathrm{T}$.

Fig. 5

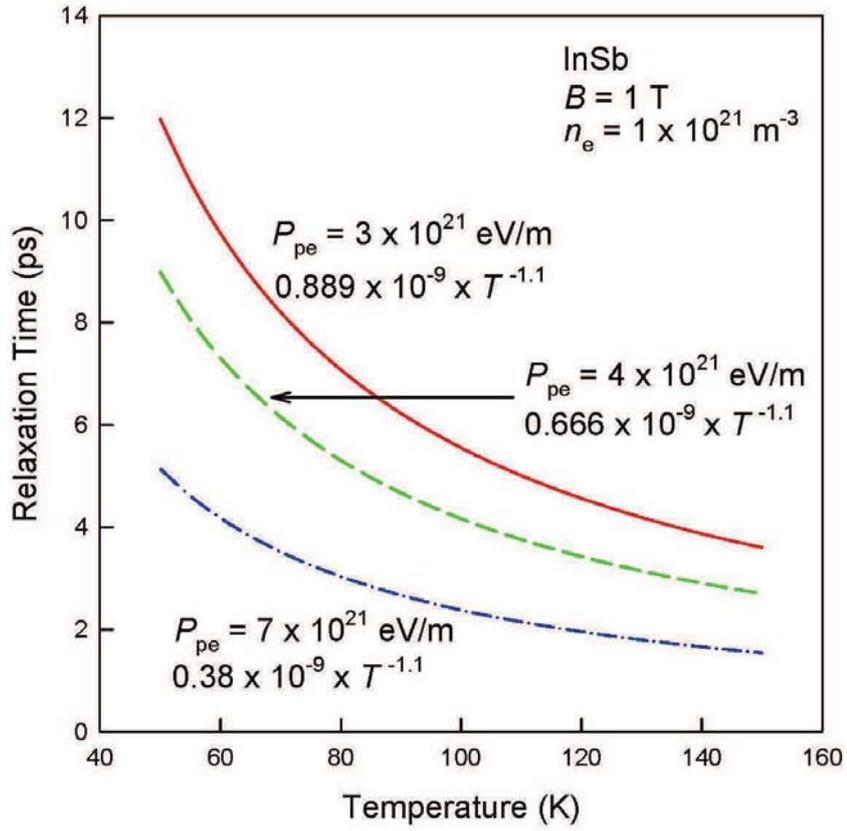

FIG. 5. Temperature dependence of the electron spin relaxation times in InSb for $n_e = 1 \times 10^{21} \mathrm{m}^{-3}$ and various $P_{pe}$'s at $B = 1\mathrm{T}$.